\documentclass[12pt,preprint]{aastex}
\usepackage{epsfig}
                                                                                
\def\vkm{km s$^{-1}$}

\def\degree{$^\circ$}
\def\arcs#1{$#1''$}
\def\arcsa#1#2{$#1^{\prime\prime}_{^\textrm{.}}#2$}

\def\solarmass{$M_\odot$}
\def\solarmasse{M_\odot}

\def\mJyb{mJy beam$^{-1}$}

\def\cmc{cm$^{-3}$}

\def\micron{$\mu$m}

\def\Ro{R_\textrm{\scriptsize 0}}

\def\H2{H$_2$}
\def\N2HP{N$_2$H$^+$}

\def\cCO{C$^{18}$O}

\def\NH3{NH$_3$}

\def\mH2{m_{\textrm{\scriptsize H}_2}}

\def\na{n_\mathrm{t}}
\def\Ro{R_\mathrm{o}}
\def\Ra{R_\mathrm{t}}

\def\Ta{T_\mathrm{t}}

\def\ha{h_\mathrm{t}}

\def\cs{c_\mathrm{s}}

\def\vp{v_\phi}

\def\iI{{\it I}}
\def\iQ{{\it Q}}
\def\iU{{\it U}}
                                                                            
\def\putfiga#1#2#3{\epsfig{scale=#1,angle=#2,figure=#3}}
\def\putfig#1#2#3{}
\def\leftblank#1{}
\def\bf#1{}

\begin{document}

\title{ALMA Dust Polarization Observations of Two Young Edge-on Protostellar
Disks}

\author{Chin-Fei Lee\altaffilmark{1,2}, Zhi-Yun Li\altaffilmark{3},
Tao-Chung Ching\altaffilmark{4}, Shih-Ping Lai\altaffilmark{4}, Haifeng
Yang\altaffilmark{3}}

\altaffiltext{1}{Academia Sinica Institute of Astronomy and Astrophysics,
P.O. Box 23-141, Taipei 106, Taiwan; cflee@asiaa.sinica.edu.tw}
\altaffiltext{2}{Graduate Institute of Astronomy and Astrophysics, National Taiwan
   University, No.  1, Sec.  4, Roosevelt Road, Taipei 10617, Taiwan}
\altaffiltext{3}{Astronomy Department, University of Virginia, Charlottesville, VA 22904}
\altaffiltext{4}{Institute of Astronomy and Department of Physics, National
Tsing Hua University, Hsinchu, Taiwan}


\begin{abstract}


{Polarized emission is detected in two young nearly edge-on protostellar
disks in 343 GHz continuum at $\sim$ 50 au ($\sim$ \arcsa{0}{12}) resolution
with Atacama Large Millimeter/submillimeter Array.  One disk is in HH 212
(Class 0) and the other in HH 111 (early Class I) protostellar system. 
Polarization fraction is $\sim$ 1\%.  The disk in HH 212 has a radius of
$\sim$ 60 au.  The emission is mainly detected from the nearside of the
disk.  The polarization orientations are almost perpendicular to the disk
major axis, consistent with either self-scattering or emission by grains
aligned with a {\it poloidal} field around the outer edge of the disk
because of optical depth effect and temperature gradient; the presence of a
poloidal field would facilitate the launching of a disk wind, for which
there is already tentative evidence in the same source.  The disk of HH 111
VLA 1 has a larger radius of $\sim$ 220 au and is thus more resolved.  The
polarization orientations are almost perpendicular to the disk major axis in
the nearside, but more along the major axis in the farside, forming roughly
half of an elliptical pattern there.  It appears that toroidal and poloidal
magnetic field may explain the polarization on the near and far side of the
disk, respectively.  However, it is also possible that the polarization is
due to self-scattering.  In addition, alignment of dust grains by radiation
flux may play a role in the farside.  Our observations reveal a diversity of
disk polarization patterns that should be taken into account in future
modeling efforts.}

\end{abstract}


\section{Introduction}

Stars generally form in magnetized molecular clouds. A fundamental problem
of star formation in magnetized clouds is the so-called ``magnetic flux
problem.'' If all of the magnetic flux that threads a typical star-forming
dense core is dragged into the central object, the stellar field strength
would be orders of magnitude stronger than the observed value
\citep{Shu1987}.  How this longstanding puzzle is resolved is unclear.  We
aim to resolve this puzzle by detecting polarized emission with Atacama
Large Millimemeter/submillimeter Array (ALMA) from two nearly edge-on {
(with an inclination of 90\degree{} for edge-on)} protostellar disks.

Specifically, we seek to determine whether some of the core magnetic flux is
dragged into the disk or not.  If yes, this flux would play a key role in
both the disk evolution and the outflow launching.  For disks that are
non-self-gravitating, their evolution is believed to be driven by the
magnetorotational instability \cite[MRI,][]{Balbus1991}.  However, in the
weakly ionized outer parts of protostellar disks, the mass accretion rate
driven by MRI is well below the observed level if the magnetic flux
threading the disk is zero \cite[e.g.,][]{Simon2013a}.  To be compatible
with observations, there must be a net magnetic flux threading the outer
parts of the disk \cite[e.g.,][]{Simon2013b}.  Therefore, the averaged
poloidal magnetic field on the disk must be non-zero.  Similarly, in one of
the two widely discussed scenarios of jet launching, the disk-wind picture
\cite[][the other being the X-wind picture, Shu et al.  2000]{Konigl2000}, a
well-ordered poloidal magnetic field is required over an extended range of
(inner) disk radii; the jet material is flung centrifugally to a high speed
along the rapidly rotating poloidal field lines like ``beads on a wire''. 
Despite the fundamental role that the poloidal magnetic field plays in both
the disk evolution and jet launching, it has never been firmly detected
before.

The reason that the poloidal magnetic field has not been detected in disks
to date through dust polarization is probably a combination of the
resolution effect, source orientation, and evolutionary stage. 
Traditionally, dust polarization is used to infer the magnetic field
morphology because the polarization can be produced by the alignment of
(spinning) dust grains with their long axes perpendicular to the magnetic
field \citep{Andersson2015}.  However, recent observations found that dust
polarization can also be produced by other two mechanisms: (1)
self-scattering from dust grains of sizes comparable to the observed
wavelength \citep{Kataoka2015,Kataoka2016a,Yang2016a,Yang2016b}, which is
expected to be more significant in older disks where the dust grains have
grown to submillimeter size, and (2) alignment of dust grains with their
short axes along the direction of radiation anisotropy
\citep{Kataoka2017,Tazaki2017}.  Thus, the search of poloidal field becomes
more complicated.

Polarization observations have been done in dust continuum with
Submillimeter Array (SMA) and Combined Array for Research in Millimeter-wave
Astronomy (CARMA) toward Class 0 sources, e.g., NGC 1333 4A
\citep{Girart2006}, HH 211 \citep{Lee2014Mag}, and a sample of others
\citep{Hull2014}.  In these sources, however, the disks remain unresolved
($< 100$ AU in diameter), and a polarization hole is seen, due to averaging
of small-scale field structures in the disks.  Once the disks are resolved,
such as the Class 0 disks of IRAS 16293B and L1527, polarization can be
detected.  However, the IRAS 16293B disk is almost face-on, only toroidal
field can be detected \citep{Rao2014}.  Although the L1527 disk is edge-on,
it is unresolved in the vertical direction, which makes it difficult to
probe the detailed disk field structure \citep{Segura2015}.  Polarization
observations have also been done toward Class I sources.  For the resolved
disk, e.g., HL Tau (evolving from Class I to Class II phase), polarization
is detected \citep{Stephens2014,Stephens2017,Kataoka2017}.  However, the
published data is more consistent with the pattern from scattering by large
grains and emission by radiative flux-aligned grains rather than from
emission by magnetically aligned grains
\citep{Yang2016a,Kataoka2016a,Stephens2017}.  { Polarization observations
have also been done toward Class II sources, but the polarization was found to
arise mainly from self-scattering by large grains \citep{Kataoka2016b}.}
Large grains are expected in relatively evolved sources that have a longer
time for the grains to grow compared to the youngest protostars.  { In
addition, self-scattering can also contribute significantly to the
polarization in disks around massive stars \citep{Fernandez2016}.}


As discussed above, to have the best chance of detecting the poloidal
magnetic field, the disks must be (1) resolved, (2) close to edge-on to
avoid depolarization due to poloidal field at a small inclination angle, and
(3) as young as possible to reduce the chance of polarization contamination
from scattering, since there would be less time for the grains to grow to
large enough sizes for efficient scattering.  These criteria are satisfied
by two bright nearby disks in Orion at a distance of $\sim$ 400 pc, one in
HH 212 (Class 0) and one in HH 111 (early Class I).  Moreover, highly
collimated jets \citep{Lee2017jet,Reipurth1999} are seen launching from
them, indicating that the disks should have poloidal field at least in the
inner part.  In addition, our ALMA data in molecular lines have resolved the
disks and found them to have a Keplerian rotation
\citep{Lee2017envdisk,Lee2016}.  In this paper, we present our polarization
observations toward these two disks to search for the poloidal field.

\section{Observations}\label{sec:obs}

Polarization observations toward the HH 111 and HH 212 protostellar systems
were carried out together with ALMA in Band 7 at $\sim$ 343 GHz in Cycle 3,
with 38-40 antennas in the Configuration C36-6.  Two executions were carried
out in 2016, one on August 30 and the other on September 3.  The projected
baselines are 14-2483 m.  The maximum recoverable size (MRS) scale is $\sim$
\arcsa{1}{4}, enough to cover the disks seen in the continuum.  One pointing
was used to map the center of the system.  The correlator was set up to have
4 continuum windows (with one at 336.5 GHz, one at 338.5 GHz, one at 348.5
GHz, and one at 350.5 GHz), with a central frequency at $\sim$ 343 GHz.  The
total time is $\sim$ 40 minutes on each HH system.


The $uv$ data were calibrated with the Common Astronomy Software
Applications (CASA) package, with quasars J5010+1800, J0552+0313 (0.31 Jy),
and J0522-3627 (3.04 Jy) as bandpass, gain, and polarization calibrators,
respectively.  A super-uniform weighting with a robust factor of 0.5 was
used for the $uv$ data to generate the continuum maps of the disks at $\sim$
343 GHz.  Notice that the $uv$ data with a distance $<$ 50 m are not well
sampled and thus excluded in our map generation.  This generates a
synthesized beam (resolution) of $\sim$ \arcsa{0}{12}.  In Stokes \iI{}
maps, the noise level is $\sim$ 0.44 \mJyb{} in HH 212 and $\sim$ 0.34
\mJyb{} in HH 111.  In Stokes \iQ{} and \iU{} maps, the noise is $\sim$ 0.03
\mJyb{} in both sources.  (Linear) Polarization fraction (degree) is defined
as $P=\sqrt{Q^2+U^2}/I$.  { According to ALMA Technical Handbook in Cycle
5, the instrumental error on $P$ is expected to be $\lesssim$ 0.2\% for the
disks of HH 212 and HH 111 VLA 1 at the phase center, because their disk
size is $\lesssim$ \arcs{1} and thus much smaller than 1/3 of the primary
beam, which is $\sim$ \arcs{17}.  It could increase to $\sim$ 0.3\% for the
disk of HH 111 VLA 2, which is located at $\sim$ \arcs{3} away from the VLA
1 source.} Polarization orientations are defined by the $E$ vectors.


\section{Results}

\subsection{HH 212}


Figure \ref{fig:HH212map}a shows the continuum map (white contours) toward
the central source in HH 212 at 343 GHz at $\sim$ \arcsa{0}{12} resolution,
with the polarized intensity map (color map) and polarization orientations
(line segments).  The continuum map shows an elongated structure
perpendicular to the jet axis.  The total flux density is $\sim$ 150$\pm$25
mJy.  As seen in Figure \ref{fig:HH212map}b, the continuum emission within
$\sim$ \arcsa{0}{2} of the central source is spatially coincident with the
dusty disk (orange map with white contours) detected before at 350 GHz at
$\sim$ \arcsa{0}{02} resolution \citep{Lee2017disk} and thus mainly traces
the dusty disk around the central source.  According to that paper, the disk
is nearly edge-on, with an inclination of $\sim$ 86\degree{} and the nearside
tilted slightly to the south.  In addition, the disk is flared with a dark
lane along the equatorial plane.  However, the disk here is mainly resolved
in the major axis and only marginally resolved in the minor axis along the
vertical direction.

\begin{figure} [!hbp]
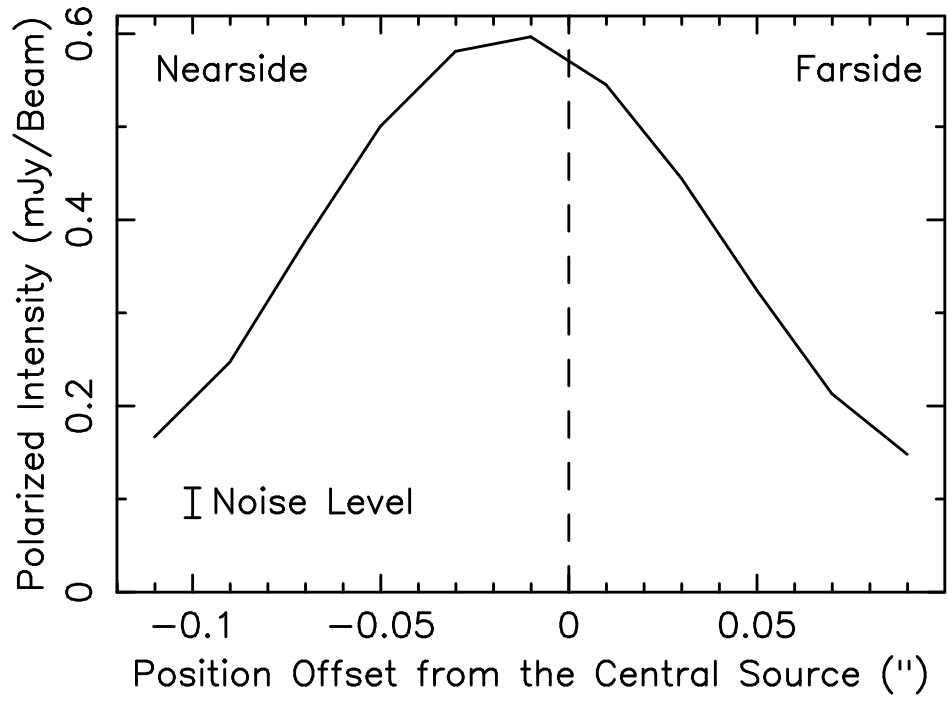

\centering
\putfiga{1}{270}{f2.eps} 
\figcaption[]
{HH 212 - Polarized intensity along the minor axis of the disk.
\label{fig:HH212Pol}}
\end{figure}

Polarized emission is detected toward the dusty disk, with a polarization
fraction of $\sim$ 1\%, as indicated by the length of the line segments. 
The polarized intensity increases toward the center.  It is slightly higher
below the major axis of the disk than above, producing a near-far side
asymmetry in the polarized intensity.  { This asymmetry is significant
because the polarized intensity in the nearside peaks at $\sim$
$-$\arcsa{0}{014} (from Gaussian fit) with a value higher than that at
\arcsa{0}{014} in the farside by more than a noise level, as shown in Figure
\ref{fig:HH212Pol}.} The polarization orientations are all roughly
perpendicular to the major axis of the disk.

\subsection{HH 111}


In HH 111, there are two sources, VLA 1 and VLA 2, with VLA 1 being the
source driving the HH 111 jet.  Figure \ref{fig:HH111map}a shows the
continuum map (white contours) toward the VLA 1 source at 343 GHz at $\sim$
\arcsa{0}{12} resolution, with the polarized intensity map (color map) and
polarization orientations (line segments).  A dusty disk has been detected
before in continuum at 230 GHz at up to $\sim$ \arcsa{0}{35} resolution
\citep{Lee2010,Lee2011,Lee2016}.  It is also close to edge-on, but with a
smaller inclination angle of $\sim$ 80\degree{} and the near side tilted to
the east \citep{Lee2016}.  This dusty disk is now better resolved at higher
resolution obtained at higher frequency.  It appears to be exactly
perpendicular to the jet axis, with a Gaussian deconvolved size of $\sim$
\arcsa{0}{54} in the major axis and $\sim$ \arcsa{0}{18} in the minor axis. 
It has a total flux density of $\sim$ 535$\pm$80 mJy.  Previously with the
\cCO{} J=2-1 line emission, the disk was found to have a Keplerian rotation
within a radius of $\sim$ \arcsa{0}{4} of the central source
\citep{Lee2016}, and this radius is roughly coincident with the edge of the
disk detected here.

Polarized emission is detected toward the disk, with a polarization fraction
of $\sim$ 1\%.  As found in HH 212, the polarized intensity increases toward
the center.  Since the disk is better resolved in the minor axis because of
its lower inclination angle and larger disk radius, the emission in the
nearside and farside can be better separated.  As can be seen, the polarized
intensity is much higher in the nearside of the disk than the farside. 
Interestingly, the polarization orientations are mainly perpendicular to the
major axis of the disk in the nearside, but bended around the jet axis in
the farside, forming the upper half of an elliptical polarization pattern. 
As a result, a polarization gap is seen right above the major axis, with
less polarization closer to the jet axis, due to the depolarization of
orthogonal polarizations there.

Figure \ref{fig:HH111map}b shows the continuum map (white contours) toward
the VLA 2 source at 343 GHz at $\sim$ \arcsa{0}{12} resolution, with the
polarized intensity map (color map) and polarization orientations (line
segments).  The VLA 2 source is located at $\sim$ \arcs{3} (1200 AU) to the
northwest of the VLA 1 source.  Its continuum emission structure is slightly
elongated at a position angle of $\sim$ 117\degree{}, with a Gaussian
deconvolved size of $\sim$ \arcsa{0}{068}, and thus can
have a major axis at this position angle.   The VLA 2 source could be driving
the HH 121 jet \citep{Sewilo2017}, because its major axis is roughly
perpendicular to the jet, which has a position angle of $\sim$
30\degree$\pm$15\degree{} (as roughly indicated by the blue and red arrows). 
As a result, the continuum emission may trace a flattened dusty envelope or a
small dusty disk around the VLA 2 source.  It has a total flux density of
$\sim$ 18.2$\pm$3.6 mJy.  The polarization orientations are perpendicular to
the major axis of the continuum emission, like that seen in HH 212.



\section{A Disk Model for HH 111} \label{sec:model}



Disk properties are needed to understand the dust polarization and then to
study if the dust polarization can be used to infer the magnetic field
morphology in the disk.  Since the disk properties of HH 212 have already
been determined by \citet{Lee2017disk}, we only aim to determine the disk
properties of HH 111 around the VLA 1 source.  Previously, a simple flat
disk model was adopted for simplicity to model the disk emission in
continuum at 230 GHz at $\sim$ \arcsa{0}{35} resolution \citep{Lee2011}. 
Since the disk is roughly resolved here in the minor axis, we adopt a flared
disk model similar to that used before to model the disk emission in HH 212
at higher resolution \citep{Lee2017disk}.  { Notice that, without having
observations in more than one frequency at higher angular resolution, we are
not able to constrain the physical properties of the disk accurately.  Thus,
the model here is really for illustrative purposes only.}


As in \citet{Lee2017disk}, we introduce a parametrized model for the disk emission, in
order to roughly obtain the properties of the disk.
For a disk in vertical hydrostatic equilibrium, the scale height $h$
in the cylindrical coordinate system is given by  
\begin{equation}
\frac{h}{R} \sim \frac{\cs}{\vp} \propto \frac{R^{-q/2}}{R^{-1/2}} = 
(\frac{R}{R_o})^{(1-q)/2}
\end{equation}
where $R$ is the cylindrical radius and $c_s$ is the isothermal sound speed
proportional to $T^{1/2}$, where $T$ is the temperature.  Assuming $T
\propto R^{-q}$, we have $c_s \propto R^{-q/2}$.  $\vp$ is the rotational
velocity assumed to be Keplerian and thus proportional to $R^{-1/2}$, as
suggested by C$^{18}$O gas kinematics \citep{Lee2016}.  In this case, the
scale height of the disk increases with the disk radius monotonically. 
However, Figure \ref{fig:HH111map}a shows that the continuum emission of the
disk becomes geometrically thinner near the outer edge, 
as seen in HH 212 \citep{Lee2017disk}.  The physical reason for the
behavior is unclear; 
self-shielding could reduce the temperature in the outer part of the disk
and thus the disk scale height \citep{Dullemond2004}.
To model the decrease of disk thickness near the outer
edge, we introduce a radius, $\Ra$, beyond which the scale height decreases
to zero at the outer radius of the disk $\Ro$.  Let $\ha$ be the scale
height of the disk at $\Ra$, we have
\begin{eqnarray}
h (R) \sim \ha  \left\{ \begin{array}{ll}
(\frac{R}{\Ra})^{1+(1-q)/2}  & \;\;\textrm{if}\;\; R < \Ra,  \\ 
\sqrt{1-(\frac{R-\Ra}{\Ro-\Ra})^2} & \;\;\textrm{if}\;\; \Ra \leq R \leq \Ro
\end{array}  \right.
\label{eq:thick}
\end{eqnarray}


The temperature is simply assumed to increase radially toward the center with
\begin{eqnarray} 
T \sim \Ta (\frac{R}{\Ra})^{-q} 
\end{eqnarray} 
where $\Ta$ is the temperature at $\Ra$ and $q$ is the power-law index.  
The number density of molecular hydrogen is assumed to be
\begin{equation}
n= \na (\frac{R}{\Ra})^{-p} \exp(-\frac{z^2}{2 h^2}) 
\end{equation} 
where $p$ is the power-law index and $\na$ is the number density in the disk
midplane at $\Ra$. Here the density is assumed to decrease 
exponentially above the disk midplane. 
Helium is included so that the mass density is $\rho=1.4 n \mH2$.

Dust opacity is needed to derive the dust emission.  For circumstellar disk,
the dust opacity formula can be given as 
\begin{equation}
\kappa_\nu = 
0.1 \left( \frac{\nu}{10^{12}\textrm{Hz}} \right) ^\beta 
\;\;\textrm{cm}^2 \;\textrm{g}^{-1}
\end{equation}
where $\beta$ is the dust opacity spectral index \citep{Beckwith1990}. 
Based on the SED (spectral energy distribution) fitting to the continuum flux around
the central source obtained at much lower resolution of a few arcseconds, 
$\beta$ was found to be $\sim$ 0.9
\citep{Lee2009HH111}.  At that low resolution, the flux comes from both the disk
and envelope.  Since the disk itself is expected to have a lower $\beta$ due to a larger grain size,
we assume $\beta \sim$ 0.6, as that found in the disk of HH 212
\citep{Lee2008}.  Thus, the dust opacity is $\kappa_\nu\sim 0.053$ cm$^2$
g$^{-1}$ at 343 GHz.  Note that the value of $\beta$ only affects the
required mass of the disk to produce the observed emission.  Smaller $\beta$
implies larger $\kappa_\nu$, and thus lower disk mass.




We used our radiative transfer code \citep{Lee2017disk} to obtain dust continuum
emission map from the model disk.  We first computed the thermal emission
from each point in the disk based on its local temperature, and then
generated a synthetic map by integrating along each line of sight the local
emission that is attenuated by the optical depth.  The disk is assumed to
have an inclination angle of $\sim$ 80\degree{}.  { The resulting map is
then convolved with the observed beam in order to generate a model map to be
compared with the observed map at the same resolution.}

\begin{figure} [!hbp]
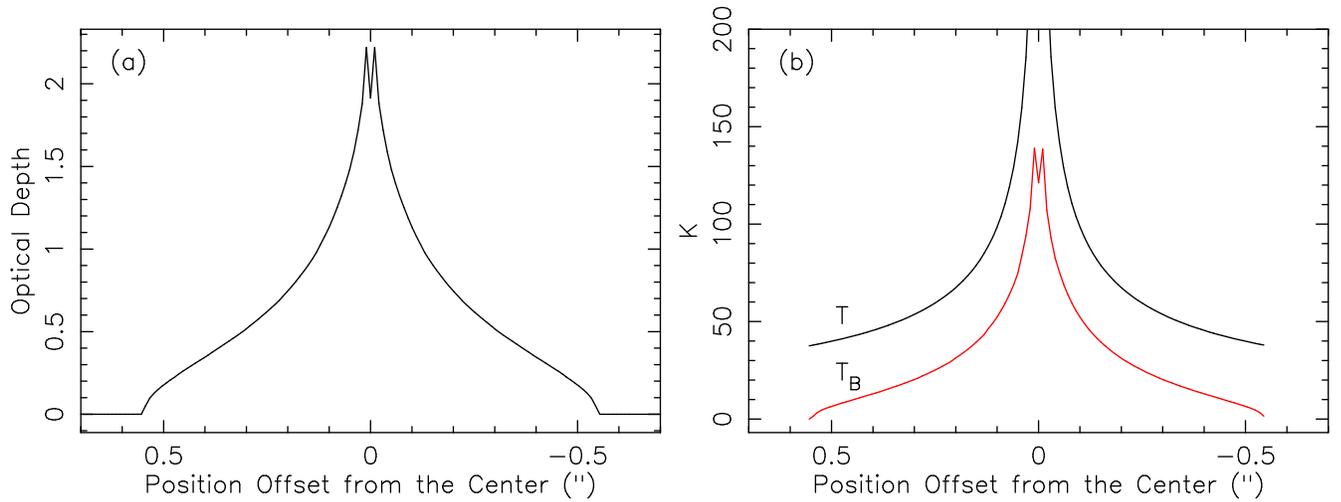

\centering
\putfiga{0.7}{270}{f5.eps} 
\figcaption[]
{Optical depth and brightness temperature in our model for the HH 111 VLA 1 disk
along the disk major axis. Panel (a) shows the optical depth.
Panel (b) shows the brightness temperature (red) in comparison 
to the temperature (black) in the model.
\label{fig:HH111ModelT}}
\end{figure}

Since HH 111 is a Class I source, we assume $q=0.5$, as that found for the
disk in the later stage \citep{Andrews2009}.  We also assume $p= 1$ and use
an eye-fitting for the rough determination of the disk properties.  The
best-fit parameters are $\Ro \sim$ \arcsa{0}{55}$\pm$\arcsa{0}{11}
(220$\pm$44 AU), $\Ra \sim$ \arcsa{0}{15}$\pm$\arcsa{0}{03} (60$\pm$12 AU),
$\ha \sim$ \arcsa{0}{08}$\pm$\arcsa{0}{02} (32$\pm$8 AU), $\na \sim
1.3\pm0.3\times10^{9}$ \cmc{}, and $\Ta\sim$ 87$\pm$17 K.  Note that the
best-fit parameters are assumed to have 20\% uncertainties.  Since $\Ra \sim
0.27\Ro$, most part of the disk (with a radius $>\Ra$) is almost flat with
roughly a constant thickness.  With $p=1$, the surface density of the disk
is roughly proportional to $r^{-1}$ for most of the disk, similar to that
found before in \citet{Lee2011} for the HH 111 disk emission at a lower
resolution and in older disks \citep{Andrews2009}.  Figures
\ref{fig:HH111Model}a and \ref{fig:HH111Model}b show the best-fit model disk
{ map on top of the observed map} and the residual, respectively.  The
residual is reasonably small, indicating that the disk model is acceptable. 
{ In this model, the dust emission is optically thin except for the
innermost \arcsa{0}{2} region, and thus the brightness temperature (before
convolved with the observed beam) is much lower than
$(1-\textrm{e}^{-\tau})$ times the temperature, as shown in Figure
\ref{fig:HH111ModelT}.}

At $R=\Ra$, the Keplerian rotation velocity is $v_\phi \sim 4.5$ \vkm{}
\citep{Lee2016}.  The mid-plane temperature there is $\sim$ 87~K, yielding
an isothermal speed of $\sim$ 0.6 \vkm{}.  Thus, the scale height at $\Ra$
is expected to be $\sim$ 8 AU, which is much smaller than the value of $\ha$
inferred from the continuum modeling.  Thus, the scale height could be
overestimated.  Radiation heating by the central protostar could help
increase the scale height but more detailed modeling is needed to quantify
this effect.  In any case, higher resolution observation to better resolve
the vertical structure is needed to confirm the disk structure.  In this
model, the disk has a total mass of \begin{equation} M_D= 1.4\,\mH2 \int
n\,2\pi\,R\,dR\,dz \sim 0.04 \,\solarmasse \end{equation} which is $\sim$
only 2\%-3\% of the mass \cite[1.3-1.8 \solarmass,][]{Lee2016} of the
central protostar, supporting that the disk can indeed rotate with a
Keplerian velocity.  { Again, since our model is based on the observation
in one single frequency, the model results here should be used with
caution.}

\section{Discussion}



As mentioned in the introduction, three mechanisms can produce the
polarization in dust emission at (sub)millimeter wavelengths: 

(1) alignment of (spinning) dust grains with their long axes perpendicular
to the magnetic field \citep{Andersson2015}, as seen in the envelope in NGC
1333 IRAS 4A \citep{Girart2006} and probably in the disk in IRAS 16293-2422
B \citep{Rao2014}.  In the envelope and disk, the temperature is expected to
increase toward the center.  In the optically thin case, the polarization
orientations will be perpendicular to the magnetic field direction, because
the emission with the polarization orientation parallel to the magnetic
field direction has a smaller opacity.  However, in the optically thick
case, the polarization orientations will be parallel to the magnetic field
direction, because the emission with the polarization orientation parallel
to the magnetic field direction can probe deeper in the envelope and disk
where the temperature (and thus the source function for emission) is higher. 
Readers can read Section 2.2 in \citet{Yang2017} for a more detailed
discussion.

(2) self-scattering from dust grains of sizes comparable to the observed
wavelength \citep{Kataoka2015,Kataoka2016a,Yang2016a,Yang2016b}, as seen in
the evolved disk in HL Tau, especially in ALMA Band 7 (0.87~mm)
\citep{Stephens2017}.  The self-scattering can be efficient in submillimeter
wavelengths when the grains have grown to bigger than $\sim$ { 100} \micron{}
\citep{Kataoka2015,Yang2017}.  For an inclined disk, it can produce a
near-far side asymmetry in polarized intensity and polarization orientations
perpendicular to the major axis \citep{Yang2017}.  This is because the
near-side of the disk surface is viewed more edge-on than the far-side. 
Larger grains and higher density will produce larger near-far side asymmetry
\citep{Yang2017}.

(3) alignment of dust grains with their short axes along the direction of
radiation flux (anisotropy) \citep{Kataoka2017,Tazaki2017}, as proposed for
the observed dust polarization detected in the evolved disk in HL Tau at 3
mm \citep{Kataoka2017}.  In the disk, the radiation flux is expected to be
radially directed, and thus the dust grains can be aligned with their short
axes in the radial direction.  As a result, the polarization orientation
would be circular around the central source for a face-on disk, elliptical
for an inclined disk, and mostly parallel to the major axis for a nearly
edge-on disk.

In the following, we discuss our observations in the context of these
mechanisms.

\subsection{HH 212}



In HH 212, the emission is mainly from the nearside of the disk, because of
the high inclination angle, large geometric thickness, and high optical
depth of the disk \citep{Lee2017disk}.  In addition, the emission detected
here is mainly from the outer edge of the disk because the emission there
has become optically thick.  { Although the disk is young in the Class 0
phase, the grain size in the outer edge of the disk may have grown quickly
to the submillimeter size \citep{Brauer2008}.  However, since the disk dust
emission is observed to be vertically extended, the grain settling is
unlikely to have taken place significantly.  Hence, the grains are likely
relatively small, although detailed modeling is required to constrain the
grain size quantitatively.  }





\begin{figure} [!hbp]
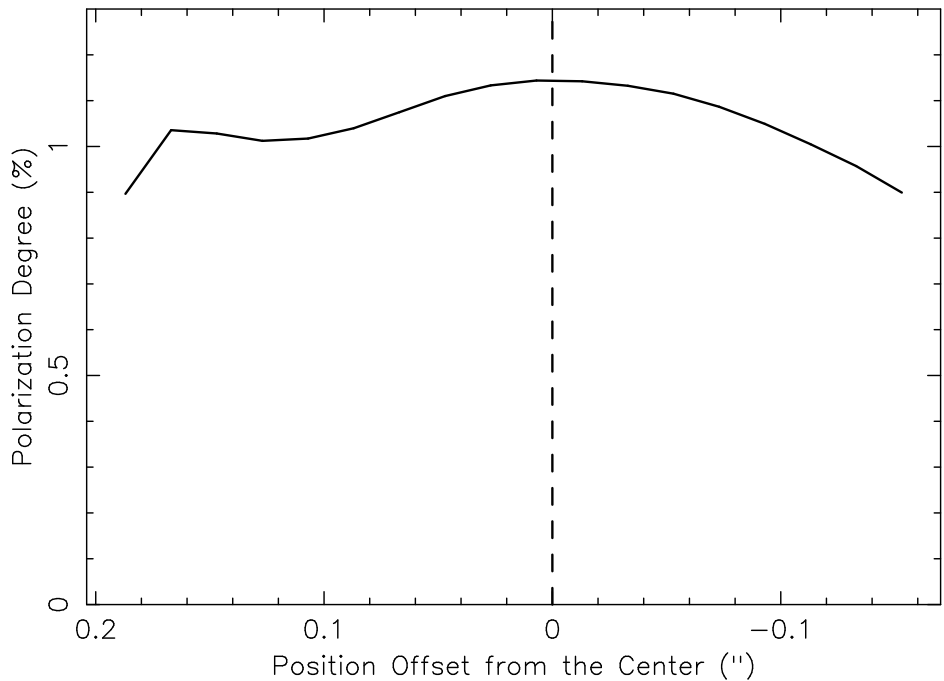

\centering
\putfiga{0.5}{270}{f6.eps} 
\figcaption[]
{Profile of the polarization degree along the major axis of the HH 212 disk,
averaged with a width of \arcsa{0}{08} perpendicular to the major axis.
\label{fig:HH212polm}}
\end{figure}

It is possible that the observed dust polarization be due to magnetic
alignment of the dust grains.  In HH 212, the temperature of the disk
increases toward the center \citep{Lee2017disk}.  Since the emission has
become optically thick in the outer edge of the disk, the inferred magnetic
field lines would be parallel to polarization orientations and thus be
mainly poloidal.  { The lack of variation in
polarization degree from the center to the disk edges
(Figure \ref{fig:HH212polm}) strengthens the case for optically thick
case somewhat, although the rather low resolution makes it inconclusive in
our view.}
If this is the case, the magnetic fields may be
responsible for launching the rotating SO/SO$_2$ outflow extending out from
the disk \citep{Tabone2017,Lee2017dwind}.  Observations at higher resolution
are needed to determine the location of the magnetic fields more accurately.



{ Since the grains may have grown to submillimeter size, it is also
possible that the observed dust polarization be due to self-scattering
\citep{Kataoka2015}.} For an inclined disk, the self-scattering tends to
produce polarization along the minor axis of the projected disk
\citep{Yang2016a,Kataoka2016a}, which is in qualitative agreement with the
pattern observed in HH 212.  In addition, for an optically thick disk such
as HH 212, the near-side of the disk is expected to be brighter in polarized
intensity than the far-side, which is also observed.  However, whether
scattering can reproduce the observations quantitatively or not remains to
be determined.  We will postpone a detailed modeling to a future
investigation.


The observed dust polarization is unlikely be due to the alignment of dust
grains by the radiation flux, because the polarization orientations are
observed to be perpendicular to the major axis, which is opposite to the
expected (parallel to the major axis) pattern.

\subsection{HH 111}


In HH 111, the disk around the VLA 1 source is more evolved than that in HH
212.  According to our simple model, it has a larger radius of $\sim$ 220 au
(\arcsa{0}{55}) and a lower inclination angle of $\sim $ 80\degree{}, and
thus the emission from the nearside and farside of the disk can be better
separated.  Interestingly, the polarization orientations are different
between the nearside and farside.  The polarized intensity shows a larger
near-far side asymmetry than that seen in HH 212.





\begin{figure} [!hbp]
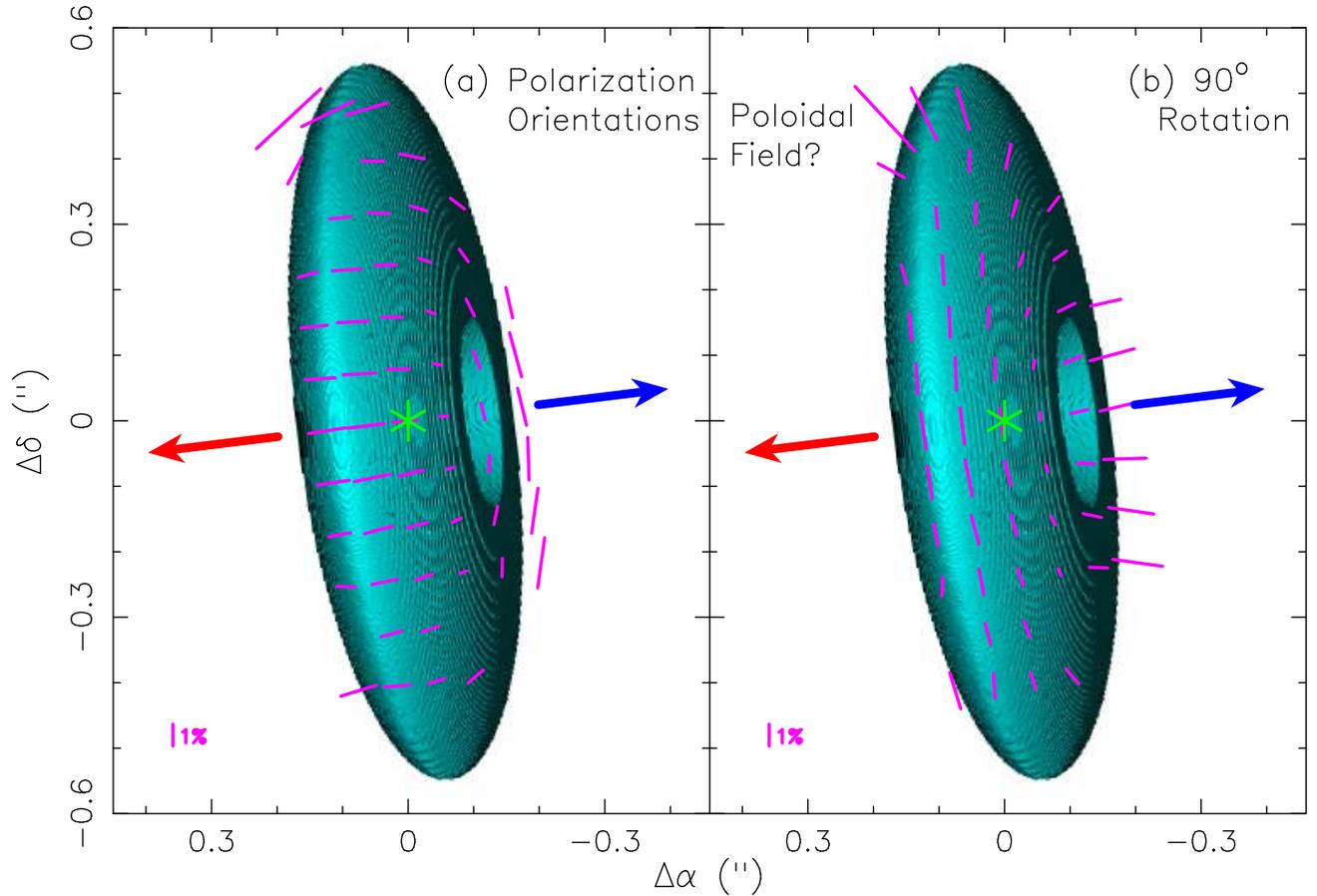

\centering
\putfiga{0.7}{270}{f7.eps} 
\figcaption[]
{ HH 111 VLA 1 --
(a) Polarization orientations plotted on the model disk structure,
{ which shows the scale height of the disk times a factor $\sqrt{2}$.}
(b) Possible magnetic field orientations, inferred by rotating the polarization orientations by 
90\degree{}, plotted on the model disk structure.
\label{fig:HH111ModelP}}
\end{figure}

Like HH 212, it is possible that the observed dust polarization of the VLA 1
disk be due to magnetic alignment of the dust grains.  Since the disk {
has a temperature increasing toward the center and is mostly optically thin
as discussed in Section \ref{sec:model}}, the inferred magnetic field
orientations could be obtained by rotating the polarization orientations by
90\degree{}.  Hence, the magnetic field in the nearside could be mainly
toroidal in the disk (see Figure \ref{fig:HH111ModelP}), as expected for a
well ionized disk where the magnetic field is expected to be wound onto a
predominantly toroidal configuration by differential rotation.  Near the
northern edge of the disk, the field lines appear to be poloidal and thus
could trace the poloidal fields dragged into the disk from the innermost
envelope (see Figure \ref{fig:HH111ModelP}b); HH 111 is a source where we have
previously uncovered evidence for a decrease of specific angular momentum in
the envelope-disk transition region, from the 1000 AU to 100 AU scale, which
is indicative of magnetic braking \citep{Lee2016}.  It is therefore not too
surprising if magnetic fields exist in the disk as well.  In the farside of
the disk above the major axis, the field lines appear to be parallel to the
jet axis, and thus could trace poloidal fields extending out from the inner
part of the disk.  As in the HH212 case, the inferred poloidal field may
also be related to the large-scale, ordered, magnetic field that is commonly
invoked for outflow driving.  However, the polarization pattern there can
also be considered as half of an elliptical pattern, and thus be
due to the dust alignment by the radiation flux in the disk
{ \citep{Tazaki2017,Kataoka2017}.}

{ It is also possible that the observed dust polarization be due to
self-scattering by dust grains.  The self-scattering can produce a near-far
side asymmetry in polarized intensity for an inclined disk reminiscent of
that observed in HH 111.  In the nearside of the disk, the observed
polarization orientations are perpendicular to the major axis, as seen in
the self-scattering model \citep{Yang2017}.  In the farside of the disk,
however, the observed polarization orientations are parallel to the major
axis, which is not seen in the self-scattering models where the disk is not
viewed as nearly edge-on \citep{Yang2017}.  It is possible that the disk has
prominent rings and gaps, as seen in HL Tau.  For such a disk,
\citet{Pohl2016} found that the polarization orientations due to
self-scattering can be parallel to the major axis, if the disk is edge-on
and optically thin.  It is plausible that a more optically thick disk with
rings and gaps that is inclined slightly away from edge-on may produce a
polarization pattern similar to that observed in HH 111 (perpendicular to
the major axis on the nearside and parallel to it on the farside).  Further
observations at higher resolution and modeling are needed to check this
possibility.}


It is more uncertain to discuss the polarization for the dust emission
around the VLA 2 source because its structure and physical properties are
poorly constrained due to insufficient resolution.  As discussed earlier,
the VLA 2 source could be driving the HH 121 jet and the continuum emission
could trace a flattened dusty envelope or a small dusty disk perpendicular
to the jet.  Assuming that the emission is optically thin, $\beta \sim 1$,
and the temperature is $\sim$ 50 K, the mass for the dust emission is only
$\sim$ 2 Jupiter mass.  The polarization along the apparent minor axis of
the disk may be due to a toroidal field (if optically thin) or
self-scattering.


\section{Conclusions}

Dust polarization has been detected toward two nearly edge-on protostellar
disks.  The interpretation of the polarization is complicated by the
presence of three different polarization mechanisms.  In HH 212, the dust
polarization is consistent with either scattering or emission by grains
aligned with a {\it poloidal} field around the outer edge of the disk
because of optical depth effect and temperature gradient.  One may be able
to tell these two apart with polarization observations at another wavelength
in the future.  For HH 111 VLA 1, it is possible that a combination of
toroidal and poloidal magnetic field may explain the polarization on the
near and far side of the disk, although we do not have good detailed models
for scattering for disks that are as edge-on as HH 111; scattering may or may
not work, and it needs more exploration.  In addition, alignment of dust
grains by radiation flux may play a role in the farside.  Perhaps, different
polarization mechanisms are operating in different parts of a disk and in
different disks.  Additional multi-wavelength polarization observations and
detailed modeling are required to make further progress in this important
field that is being revolutionized by ALMA.

\acknowledgements


{ We thank the anonymous referee for useful comments.} 
C.-F.L. thanks A. Kataoka and T. Hoang for fruitful discussion.  This paper
makes use of the following ALMA data: ADS/JAO.ALMA\# 2015.1.00037.S.  ALMA
is a partnership of ESO (representing its member states), NSF (USA) and NINS
(Japan), together with NRC (Canada), MoST and ASIAA (Taiwan), and KASI
(Republic of Korea), in cooperation with the Republic of Chile.  The Joint
ALMA Observatory is operated by ESO, AUI/NRAO and NAOJ.  C.-F.L. 
acknowledges grants from the Ministry of Science and Technology of Taiwan
(MoST 104-2119-M-001-015-MY3) and the Academia Sinica (Career Development
Award).  ZYL is supported in part by NASA NNX14AB38G and NSF AST-1313083 and
1716259.




\def\nat{Natur}

\begin{figure} [!hbp]
\centering
\putfiga{1}{270}{f1.eps} 
\figcaption[]
{HH 212 -
(a) The continuum map (white contours) toward
the central source (asterisk) at 343 GHz at $\sim$ \arcsa{0}{12} resolution,
with the polarized intensity map (color map) and polarization orientations
(line segments).  { The blue and red arrows show the approaching
and receding sides of the jet axis, respectively.}
The contours start at 6$\sigma$ with a step of 10$\sigma$, where $\sigma=0.44$ \mJyb{}. 
(b) The same polarization orientations 
on the continuum map of the dusty disk (orange map with white contours)
obtained before at \arcsa{0}{02} resolution \cite[see Figure 1d in][]{Lee2017disk}.
The disk on the side of the redshifted jet is the near side.
\label{fig:HH212map}}
\end{figure}

\begin{figure} [!hbp]
\centering
\putfiga{0.7}{270}{f3.eps} 
\figcaption[]
{HH 111 -
(a) The continuum map (white contours) toward the VLA 1 source (asterisk) at
343 GHz at $\sim$ \arcsa{0}{12} resolution, with the polarized intensity map
(color map) and polarization orientations (line segments).  
{ The blue and red arrows show the approaching
and receding sides of the jet axis, respectively.}
The contours
start at 6$\sigma$ with a step of 15$\sigma$, where $\sigma=0.34$ \mJyb{}. 
(b) Same as (a) but toward the VLA 2 source (asterisk).  The blue and red
arrows show roughly the mean orientation of the HH 121 jet, which has a
position angle of 30\degree$\pm$15\degree{} \citep{Sewilo2017}.  The
contours start at 4$\sigma$ with a step of 10$\sigma$, where $\sigma=0.34$
\mJyb{}.
The disk on the side of the redshifted jet is the near side.
\label{fig:HH111map}}
\end{figure}

\begin{figure} [!hbp]
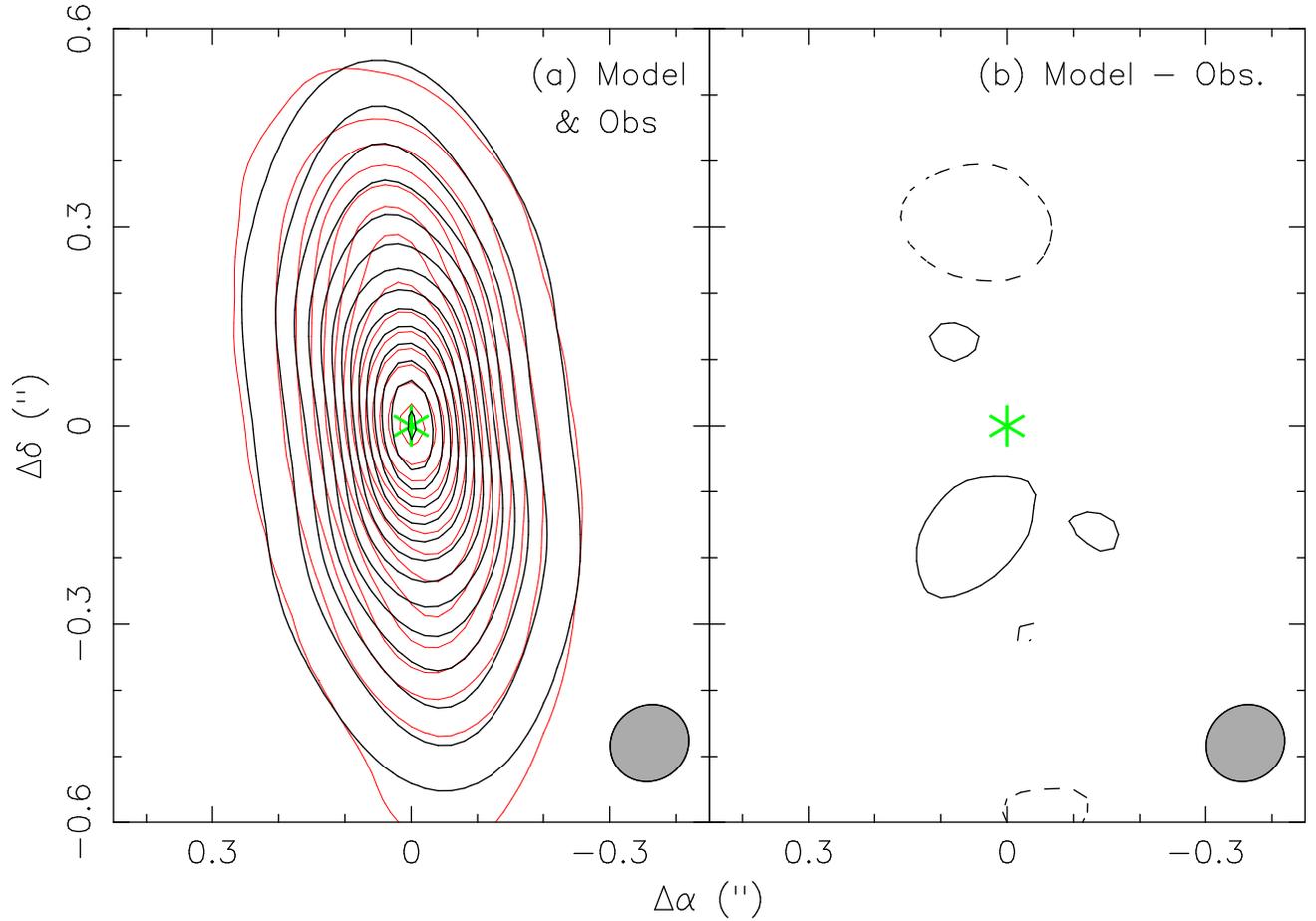

\centering
\putfiga{0.7}{270}{f4.eps} 
\figcaption[]
{Modeling of the HH 111 VLA 1 disk. (a) { Model disk emission map (black contours)
on top of the observed map (red contours).} (b) Residual (Model - Observation).
In both panels, the contours start at the same level with the same step as in Figure \ref{fig:HH111map}a.
\label{fig:HH111Model}}
\end{figure}

\end{document}